\pdfoutput=1
\documentclass[runningheads]{llncs}
\usepackage[T1]{fontenc}
\usepackage{amsfonts,amssymb}
\usepackage[ruled,noline]{algorithm2e}
\usepackage{lineno}
\usepackage{floatrow}
\usepackage{hyperref}
\usepackage{float}
\usepackage{multirow}
\usepackage{eurosym}
\usepackage{inputenc}
\usepackage{amsmath}
\usepackage{booktabs}
\usepackage[misc,geometry]{ifsym}
\usepackage{graphicx}
\usepackage{color}

\begin{document}

\title{DiffMotion: Speech-Driven Gesture Synthesis Using Denoising Diffusion Model
}

\titlerunning{DiffMotion: Diffusion-based Speech-Driven Gesture Synthesis Model}

\author{Fan Zhang\inst{1,2}\orcidID{0000-0002-9534-1777} 
\and Naye Ji\inst{2}(\Letter)\orcidID{0000-0002-6986-3766} 
\and Fuxing Gao\inst{2} 
\and Yongping Li\inst{1,3}}

\authorrunning{F. Zhang et al.}

\institute{Faculty of Humanities and Arts, Macau University of Science and Technology, Macau, China 
\and College of Media Engineering, Communication University of Zhejiang, China
\email{\{fanzhang,jinaye,fuxing\}@cuz.edu.cn}
\and College of Digital Technology and Engineering, Ningbo University of Finance \& Economics \\ \email{liyongping@nbufe.edu.cn}}

\maketitle              

\begin{figure}[H]
\includegraphics[width=0.8\textwidth]{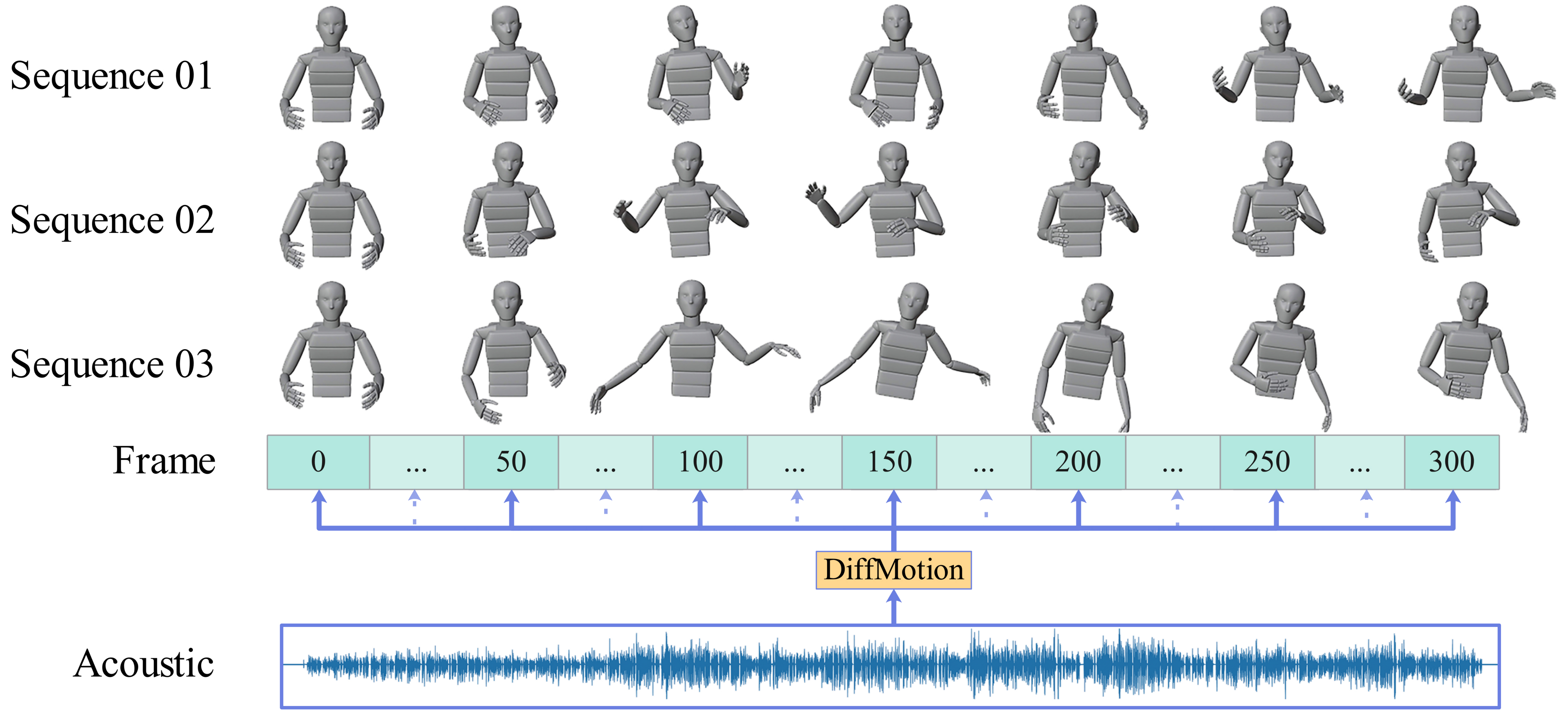}
\begin{center}
\caption{Random samples from DiffMotion can give many distinct yet natural output gestures within and between sequences, even if the input speech audio is the same.}
\end{center}\label{fig:result showcase} 
\end{figure}

\begin{abstract}
Speech-driven gesture synthesis is a field of growing interest in virtual human creation. However, a critical challenge is the inherent intricate one-to-many mapping between speech and gestures. Previous studies have explored and achieved significant progress with generative models. Notwithstanding, most synthetic gestures are still vastly less natural. This paper presents \emph{DiffMotion}, a novel speech-driven gesture synthesis architecture based on diffusion models. The model comprises an autoregressive temporal encoder and a denoising diffusion probability Module. The encoder extracts the temporal context of the speech input and historical gestures. The diffusion module learns a parameterized Markov chain to gradually convert a simple distribution into a complex distribution and generates the gestures according to the accompanied speech. Compared with baselines, objective and subjective evaluations confirm that our approach can produce natural and diverse gesticulation and demonstrate the benefits of diffusion-based models on speech-driven gesture synthesis. Project page:  \href{https://zf223669.github.io/DiffMotionWebsite/} {https://zf223669.github.io/DiffMotionWebsite/.}
\end{abstract}

\section{Introduction}
Recently, 3D virtual human technology has become increasingly popular with the rise of the metaverse. To provide natural and relatable characters, one main task is to implement non-verbal(co-speech) gestures that look natural and match the verbal ones that communicate like humans. To date, despite motion capture systems meeting this task, they require particular hardware, space, and actors, which can be expensive. Automatic generation is the cheapest way to generate gestures, which does not require human effort at production time. Speech-driven gesture generation is one considerable potential solution. However, the primary challenge for generating relevant and well-timed gestures from input speech is the inherent cross-modal one-to-many mapping between speech and gesture. That is, it is questioning to model the connection, which is that the same utterance is often accompanied by significantly different gestures at different times, even for the same or other speakers\cite{brand_matthew_voice_1999}.

Under the assumption of one-to-one mapping, previous rule-based or deterministic deep learning methods fail to achieve this task. The former limited to the provided gesture units results in repetitive movements, and the latter, trained by minimizing a mean square error, is prone to mean pose. Thus, the present research has shifted to probabilistic generative models(such as GANs, VAEs, and Normalizing Flow). Despite that, most synthetic gestures are still significantly less natural compared with the original motion-capture dataset\cite{yoonGENEAChallenge20222022}. Diffusion models which can generate high-quality and diverse samples have shown tremendously impressive results on various generation tasks. Nevertheless, diffusion models have yet gained little attention in speech-driven gesture synthesis tasks. 

This paper proposes \textbf{DiffMotion}, a novel diffusion-based probabilistic architecture for speech-driven gesture generation. The model learns on sizable sets of unstructured gesture data with zero manual annotation. Furthermore, as shown in Fig.\ref{fig:result showcase}, our method can estimate natural, various gestures, even those not present in the dataset. Our contributions are as follows:
\begin{enumerate}
\item [(1)] We propose DiffMotion, the first instance of the Diffusion-based generative model, to solve the cross-modal speech-driven gesture synthesis.

\item [(2)] We innovatively integrated an Autoregressive Temporal Encoder and a Denoising Diffusion Probabilistic Module, which can learn the complex regularities between gestures and the accompanying speech and generate realistic, various motions which match the rhythm of the speech.

\item [(3)] Experiments show that DiffMotion outperforms state-of-the-art baselines, objectively and subjectively.

\end{enumerate}
\section{Related Work}
Due to the research now shifting from \emph{deterministic} to \emph{generative} model, we only discuss the novel approaches for speech-driven gesture generation briefly.

Ylva et al.\cite{ferstl_ylva_multi-objective_2019} introduced GANs\cite{goodfellowianGenerativeAdversarialNets2014} for converting speech to 3D gesture motion using multiple discriminators. Though this approach improves significantly concerning standard regression loss training, the dataset needs to be hand-annotated. Further, they admitted the generated motions lacked realism and discontinuity because they assumed it was a gesture phase classification task. Impressively, owning to the probabilistic generative models, called \emph{Normalizing Flows}\cite{dinhNiceNonlinearIndependent2014a}\cite{dinhDensityEstimationUsing2016a}, which can tackle the exact log-likelihood and latent-variable inference, Alexanderson et al. \cite{alexanderson_simon_style-controllable_2020}constructed a Glow-based\cite{henter_gustav_eje_moglow_2020,kingmadiederikp.GlowGenerativeFlow2018} network derived from normalizing flows and RNN that successfully modeled the conditional probability distribution of the gestures given speech as input and obtained various motions given the same speech signal. Further on, Taylor et al. \cite{taylorsarahSpeechDrivenConversationalAgents2021} extended normalizing flows by combining variational autoencoder, demonstrating that the approach can produce expressive body motion close to the ground truth using a fraction of the trainable parameters. Though normalizing flows are powerful enough to capture high-dimensional complexity yet still trainable, these methods require imposing topological constraints on the transformation\cite{dinhNiceNonlinearIndependent2014a,dinhDensityEstimationUsing2016a,zhangDiffusionNormalizingFlow2021}.     

\emph{Diffusion} models(A survey in \cite{yangDiffusionModelsComprehensive2022}), the more flexible architectures, use parameterized Markov chain to convert as simple distribution into complex data distribution gradually and can be efficiently trained by optimizing the Variational Lower Bound. After overtaking GAN on image synthesis\cite{dhariwalDiffusionModelsBeat2021a,ho_denoising_2020}, the diffusion model has shown remarkable impressive results on various generation tasks, such as computer vision\cite{liSrdiffSingleImage2022}, and natural language processing\cite{austinStructuredDenoisingDiffusion2021}. In particular, the diffusion model has demonstrated impressive results in multi-modal modeling\cite{avrahamiBlendedDiffusionTextdriven2022,zhuDiscreteContrastiveDiffusion2022}, and time series forecasting \cite{rasul_multivariate_2020}, which owns the enormous prospect in cross-model sequence-to-sequence modeling and generation. However, it has yet earned little attention in speech-driven gesture synthesis tasks. Inspired by the discussions above, we design to construct a diffusion-based architecture to explore the capability of the diffusion model in speech-driven gesture generation.

\section{Our Approach} \label{sec:Our Approach}

We present  \textbf{DiffMotion} for cross-modal speech-driven gesture synthesis. The model aims to generate gesticulations conditioned on speech features and historical gestures. In this section, we first formulate the problem in Sec.\ref{sec:problemformulation} and then elaborate on the DiffMotion architecture in Sec.\ref{sec:architecture}. Finally, the training and inference process is described in Sec.\ref{sec:training} and Sec.\ref{sec:inference}.

\subsection{Problem Formulation}\label{sec:problemformulation}
We denote the gesture features and the acoustic signal as \(x^0_t \in [x^0_1,...,x^0_t,...,x^0_T]\) and \(c_t\in[c_1,...,c_t,...,c_T]\), where  \(x^0_t = \mathbb{R}^D\) is 3D skeleton joints angle at frame \(t\), and \(D\) indicates the number of channels of the skeleton joints. \(c_t\) is the current acoustic subsequence signal constructed by the sub-sequence excerpted from speech acoustic feature sequence \(\left[a_1,...,a_T\right]\), and \(T\) is the sequence length.  Let  \(p_\theta(\cdot)\) denote the Probability Density Function(PDF), which aims to approximate the actual gesture data distribution  \(p(\cdot)\) and allows for easy sampling. Given \(t^{\prime}\) as the past frame before current frame \(t\), and \(\tau\) as the past window frames and the initial motion \(x^0_{1:\tau}\), we are tasked with generating the pose \(x^0_t \sim p_\theta(\cdot)\) frame by frame according to its conditional probability distribution given historical poses \( x^0_{t^{\prime}-\tau:t^{\prime}-1}\) and acoustic signal \(c_t\)  as covariate:
\begin{equation}
\begin{aligned}
x_t^0 \sim p_\theta\left(x^0_t|x^0_{t-\tau:t-1},c_t\right) &\approx p(\cdot) := p\left(x^0_t|x^0_{t-\tau:t-1},c_t\right) \\&:=  p\left(x^0_{1:\tau}\right)  \cdot \prod\nolimits_{t^{\prime}=\tau+1}^{t} p\left(x^0_{t^{\prime}} | x^0_{t^{\prime}-\tau:t^{\prime}-1}, c_{t^{\prime}}\right)
\end{aligned}\label{eq:qxc}
\end{equation}

The autoregressive temporal encoder extracts the conditional information, and the \(p_\theta(\cdot)\) aims to approximate \(p(\cdot)\) that is trained by denoising diffusion module. We discuss these two modules in detail in Sec.\ref{sec:architecture}.
\subsection{DiffMotion Architecture}\label{sec:architecture}

DiffMotion architecture consists of two modules: Autoregressive Temporal Encoder (AT-Encoder) and Denoising Diffusion Probabilistic Module(DDPM). The whole architecture is shown in Fig.\ref{fig:schematic}.

\subsubsection{Autoregressive Temporal Encoder(AT-Encoder).}  Multi-layer LSTM is adopted for AT-Encoder to encode the temporal context of speech acoustic features and past poses up to frame \(t-1\) via updated hidden state \(h_{t-1}\) : 
\begin{equation}\label{eq:ht}
h_{t-1} = g\left(x^0_{t-\tau:t-1},c_t,h_{t-2}\right) = \rm LSTM_\theta\left(concatenate\left(x^0_{t-\tau:t-1},c_t\right),h_{t-2}\right),
\end{equation}
where \(h_t\) represents the LSTM hidden state evolution.  \(\rm LSTM_\theta\) is parameterized by sharing weights \(\theta\) and \(h_0 = 0\). Then, we use a sequence of neutral(mean) poses for the initial motion \(x_{1:\tau}\). Thus we can approximate Eq.(\ref{eq:qxc}) by the \emph{Motion Diffusion} model, and its negative log-likelihood(NLL) is :
\begin{equation}\label{eq:DMobjective}
\begin{aligned}
p_\theta\left(x^0_{1:\tau}\right)\cdot\prod\nolimits_{t^{\prime}=\tau+1}^{t} p_\theta\left(x_{t^{\prime}}^0| h_{t^{\prime}-1}\right), \quad
NLL:= \Sigma_{t^{\prime}=\tau+1}^{t} -\log p_\theta\left(x_{t^{\prime}}^0 | h_{t^{\prime}-1}\right)
\end{aligned}
\end{equation}
\subsubsection{Denoising Diffusion Probabilistic Module(DDPM).}The DDPM is a latent variable model\cite{sohl-dickstein_deep_2015} of the form $p_\theta :=  \smallint p_\theta \left(x^{0:N}\right)dx^{1:N}$, where $x^1,...,x^N$ are latent of the same dimensionality as the data $x^n$ at the \(n\)-th diffusion time stage.  The module contains two processes, namely \emph{diffusion process}  and \emph{generation process}. At training time, the diffusion process gradually converts the original data(\(x^0\)) to white noise(\(x^N\)) by optimizing a variational bound on the data likelihood. At inference time, the generation process recovers the data by reversing this noising process through the Markov chain using Langevin sampling\cite{paul_sur_1908}. The gesture can be generated by sampling from the conditional data distribution at each frame and then fed back to the AT-Encoder for producing the next frame. The Markov chains in the diffusion process and the generation process are:
\begin{equation}
\begin{aligned}
&p\left(x^n|x^0\right) = \mathcal{N}\left(x^n; \sqrt{\overline{\alpha}^n} x^0, \left(1-\overline{\alpha}^n\right)I\right)   \quad and\\ 
&p_\theta\left(x^{n-1}|x^n, x^0\right) = \mathcal{N}\left(x^{n-1}; \tilde{\mu}^n\left(x^n, x^0\right), \tilde{\beta}^n I\right),
\end{aligned}
\label{eq:cumulativeProduct}
\end{equation}
where \(\alpha^n := 1 - \beta^n\) and \(\overline{\alpha}^n := \prod_{i=1}^n \alpha^i\). As shown by \cite{ho_denoising_2020}, \(\beta^n\) is a increasing variance schedule \(\beta^1,...,\beta^N\) with \(\beta^n \in (0,1)\), and \(\tilde{\beta}^n := \frac{1-\overline{\alpha}^{n-1}}{1-\overline{\alpha}^n}\beta^n\).

\begin{figure}[htp]
    \includegraphics[width=0.6\textwidth]{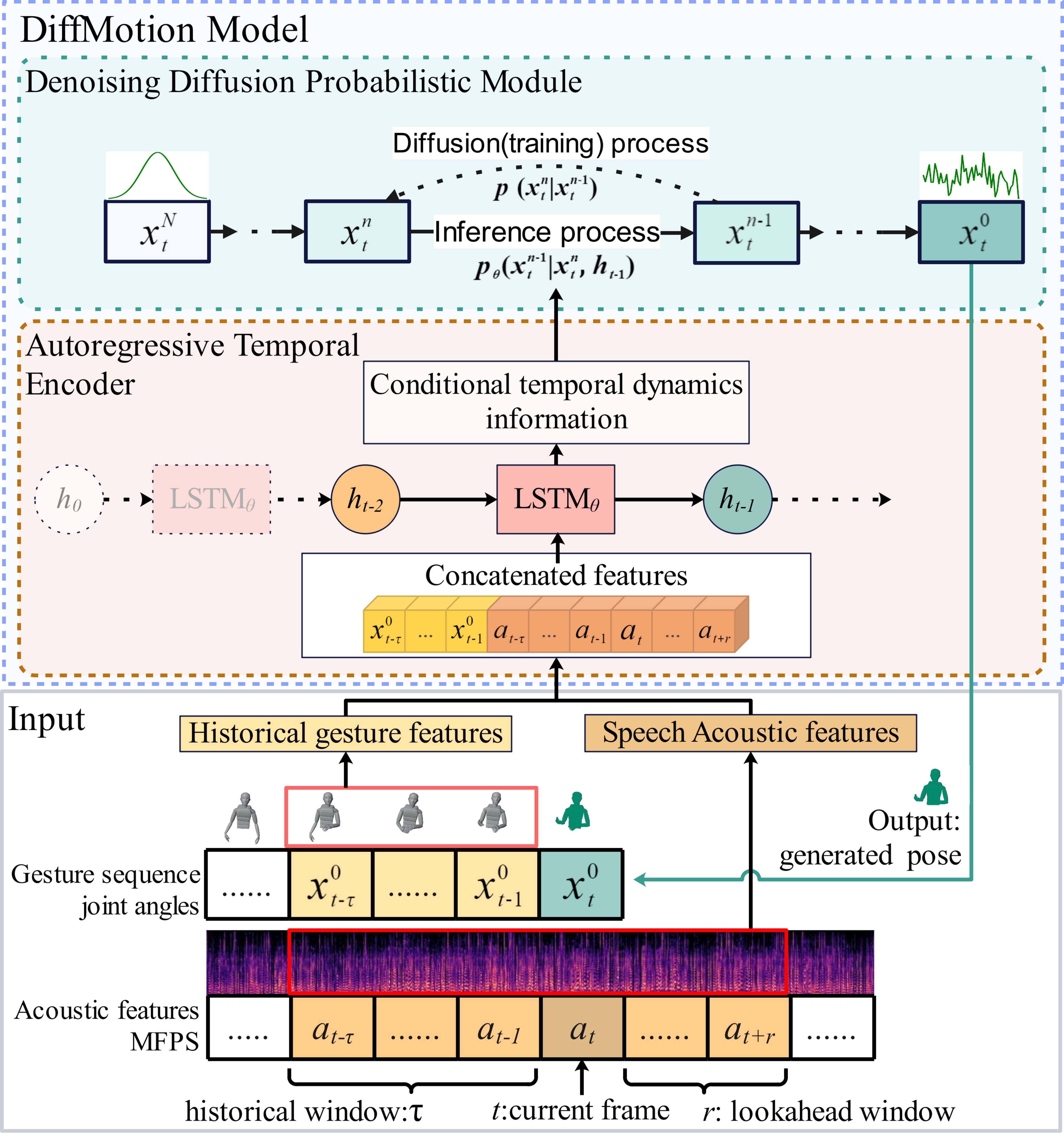}
    \caption{DiffMotion schematic. The model consists of an Autoregressive Temporal Encoder and a Denoising Diffusion Probabilistic Module.}
    \label{fig:schematic}
\end{figure}

\subsection{Training}\label{sec:training}
The training objective is to optimize the parameters \(\theta\) that minimize the NLL via Mean Squared Error(MSE) loss between the true noise \(\epsilon\sim\mathcal{N}\left(0,I\right)\)  and the predicted noise \(\epsilon_\theta\):

\begin{equation}
\label{eq:objective2}
\mathbb{E}_{x^0_t, \epsilon, n}[||\epsilon - \epsilon_\theta\left(\sqrt{\overline{\alpha}^n x^0_t}+\sqrt{1-\overline{\alpha}^n}\epsilon , h_{t-1},n\right)||^2],
\end{equation}  Here \(\epsilon_\theta\) is a neural network, which uses input \(x_t^0\) , \(h_{t-1}\) and \(n\) that to predict the \(\epsilon\), and contains the similar architecture employed in \cite{rasul_autoregressive_2021}. The complete training procedure is outlined in Algorithm \ref{alg:Training}.

\begin{algorithm}[H]
    \caption{Training for each frame  $t  \in [\tau+1,T]$}
    \KwIn{data $x^0_t \sim p\left(x^0_t|x^0_{t-\tau:t-1},c_t\right)$ and LSTM state $h_{t-1}$}
    \Repeat{converged}{Initialize $n \sim$ Uniform$(1,...,N)$ and $\epsilon \sim \mathcal{N}(0,I)$
    \\Take gradient step on
    \\ $$\nabla_\theta||\epsilon-\epsilon_\theta\left(\sqrt{\overline{\alpha}_n}x^0_t+\sqrt{1-\overline{\alpha}_n}\epsilon, h_{t-1},n\right)||^2$$}    \label{alg:Training}
\end{algorithm}
\subsection{Inference}\label{sec:inference}
After training, we expect to use variational inference to generate new gestures matching the original data distribution(\(x_t^0 \sim p_\theta\left(x^0_t|x^0_{t-\tau:t-1},c_t\right)\)). Firstly, we run the AT-Encoder over the sequence of neutral poses for the initial motion \(x_{1:\tau}\) to obtain the hidden state \(h_{t-1}\) via Eq.\ref{eq:ht}. Then we follow the sampling procedure in Algorithm  \ref{alg:Inference} to obtain a sample \(x^0_{t}\) of the current frame. The \(\sigma_\theta\) is the standard deviation of the \(p_\theta \left(x^{n-1}|x^n\right)\). We choose  \(\sigma_\theta := \tilde{\beta}^n\). 

\begin{algorithm}
\label{alg:Inference}
\SetKwFor{For}{for}{do}{end\enspace for}
\SetKwIF{If}{ElseIf}{Else}{if}{then}{else if}{else}{end\enspace if}
\SetKw{Return}{Return:}
\caption{Sampling $x_t^0$ via annealed Langevin dynamics}
\KwIn{ noise $x_t^N \sim \mathcal{N}(0,I)$ and state $h_{t-1}$} 
\For {$n = N$ \emph{\KwTo} $1$}{
	\eIf{$n>1$}{$z \sim \mathcal{N}(0,I)$}{$z = 0$}
	$x_t^{n-1}=\frac{1}{\sqrt{\alpha^n}}\left(x_t^n - \frac{\beta^n}{\sqrt{1-\overline{\alpha}^n}}\epsilon_\theta\left(x_t^n,h_{t-1},n\right)\right)+\sqrt{\sigma_\theta}z$
}
\Return{$x^0_t$}
\end{algorithm}

During inferencing, the past poses \([x^0_{t-\tau-1},...,x^0_{t-1}]\) and acoustic features \([a_{t-\tau},...a_{t+r}]\) are concatenated and sent to the AT-Encoder for extract the context, then as a conditional information to Diffusion Module to generate current gesture(\(x_t^0\)).  DiffMotion outputs one gesture in each frame and is then fed back to the generated gesture sequence. At the same time, the past pose window slides from \([t-\tau, t-1]\) to  \([t-\tau+1, t]\), and the acoustic window also moved forward by one frame for the next gesture(\(x_{t+1}^0\)) generation, as shown in Fig.\ref{fig:schematic}.

\section{Experiments}
We compare DiffMotion(\textbf{DM}) with previous baselines objectively and subjectively. For fair and comparable, we select baselines followed by: 1) using Trinity Gesture Dataset\cite{ferstl_ylva_investigating_2018} recommended by GENEA Workshop\cite{kucherenko2020genea}; 2) Skeleton structure is consistent with DM; 3) Joint angles represented by exponential map\cite{grassia_f_sebastian_practical_1998}; 4) Open source provided. A flow-based method, called StyleGestures(\textbf{SG})\cite{alexanderson_simon_style-controllable_2020}, meets the requirement above. Meanwhile, the Audio2Getsture method (\textbf{AG})\cite{lijingAudio2GesturesGeneratingDiverse2021} was introduced for only subjective evaluation since its output format is somewhat different from \textbf{DM} and \textbf{SG}. All experiments include the ground truth(\textbf{GT}). We focus on 3D upper body beat gesture, which makes up more than 50\% of all co-speech gestures and is rhythmically connected to the accompanying speech\cite{mcneill_hand_1992}.

\subsection{Training-data Processing}
Trinity Gesture Dataset we train on includes 23 takes, totaling 244 minutes of motion capture and audio of a male native English speaker producing spontaneous speech on different topics. The actor's motion was captured with 20 Vicon cameras at 59.94 frames per second(fps), and the skeleton includes 69 joints.

The upper-body skeleton was selected from the first spine joint to the hands and head and excluded the finger motion, keeping only 15 upper-body joints. We followed the data process method presented by \cite{alexanderson_simon_style-controllable_2020} and obtained 20,665×2 samples. Each with 80×27 speech features(80 frames(4s) with 27-channel Mel-frequency power spectrograms, MFPS) as input and 80×45 joint angle features as output. The frame per second is downsampled from 60 fps to 20 fps. Each joint angle was represented by an exponential map to avoid discontinuities. 
\subsection{Model Settings}
The LSTM in AT-Encoder consists of 2 layers with hidden state \(h_t \in \mathbb{R}^{512}\). The similar network set, proposed by \cite{rasul_multivariate_2020}, is employed for  \(\epsilon_\theta\). 
The quaternary  variance schedule  starts from \(\beta_1=1\times 10^{-4}\) till \(\beta^N=0.1\). We set quantile=0.5 at the inference stage for catching the median value.

In truth, the gestures must be prepared in advance \cite{mcneill_david_gesture_2008,kendon_gesticulation_1980}. We let the control inputs $c_t$ at time instance $t$ contain not only the current acoustic features $a_t$ but also the window of surrounding $c_t=a_{t-\tau:t+r}$, where the lookahead $r=20$ is set so that a sufficient amount of future information can be taken into account, shown in Fig.\ref{fig:schematic}. To avoid animation jitter, we apply Savitzky-Golay Smoothing Filter\cite{pressSavitzkyGolaySmoothingFilters1990} to filter all joint channels of the sequence along the frame and set the window length to 31 and poly order to 4.

The model is built on TorchLightning framework using a batch size of 80, Adam optimizer with a learning rate of \(1.5\times 10^{-3}\). All experiments run on an Intel i9 processor and a single NVIDIA GTX 3090 GPU.
\subsection{Objective Evaluation}
We quantitatively compare DiffMotion \textbf{DM} with  \textbf{SG} and the \textbf{GT} in the team of realism, Time consistency, and diversity: 1) \textbf{Realism}: We adopt \(L_1\) distance of joint position\cite{lijingAudio2GesturesGeneratingDiverse2021} and the Percentage of Correct 3D Keypoints(PCK)\cite{yangyiArticulatedHumanDetection2012} to evaluate the realism of the generated motion. 2) \textbf{Time consistency}: A Beat Consistency  Score(BCS) metric\cite{lijingAudio2GesturesGeneratingDiverse2021} is introduced to measure the motion-speech beat correlation(time consistency). 3) \textbf{Diversity}: The diversity metric evaluates the variations among generated gestures and in a sequence. We synthesize multiple motion sequences sampled N times for the same speech input and then split each sequence into equal-length clips without overlap. Finally, we calculate the averaged \(L_1\) distance of the whole clips. 

The results are listed in Table \ref{tab:QuantitativeEvaluation}.  The quantitative results show that our method outperforms \textbf{SG}  on realism and time consistency. On the Diversity metric, our model achieves higher score(\(6.25\) on average) than \textbf{SG}(\(3.3\)) and \textbf{GT}(\(2.6\)). It indicates that both \textbf{DM} and \textbf{SG} have the capability to generate novel gestures that are not presented in \textbf{GT}, and our model obtains a better result than \textbf{SG},  due to the Diffusion Model can sample a wider range of samples\cite{rasul_multivariate_2020}. 

\floatsetup[table]{capposition=top}
\begin{table}
\caption{Quantitative results. ↑ means higher is better, ↓ means lower is better. We perform 20 tests and report their average and best scores(in parentheses).}\label{tab:QuantitativeEvaluation}
\renewcommand{\belowrulesep}{0pt} 
\renewcommand{\aboverulesep}{0pt}
\begin{tabular}{@{}c|c|c|c|c@{}}
\toprule
Method   & L\_1↓                                 & PCK↑                                 & BCS↑                           & Diverisity↑ \\ \midrule
DM(Ours) & \textbf{11.6(10.26)} & \textbf{0.61(0.35)} & 0.79(0.80)                     & \textbf{6.25(6.31)}  \\
SG       & 16.35(15.22)                          & 0.41(0.50)                           & 0.68(0.69)                     & 3.3(3.6)    \\
GT       & 0                                     & 0                                    & \textbf{0.93} & 2.6         \\ \bottomrule
\end{tabular}
\end{table}

\begin{table}[htp]
\caption{Parameter counts, training time, and average synthesis time per frame with 95\% confidence intervals.}\label{tab:Parameters}
\renewcommand{\belowrulesep}{0pt} 
\renewcommand{\aboverulesep}{0pt}
\begin{tabular}{c|c|c|c}
\toprule
Method                  & Param.Count↓                   & Train.time↓                           & Synth.time↓                                                    \\ \midrule
DM(Ours)                & \textbf{10.4M}                 &\textbf{8.02min}                            & 1.60±0.06s                                            \\
SG                      & 109.34M                        &6.2H                            & \textbf{0.08±0.06s} \\ \bottomrule                                                  
\end{tabular}
\end{table}

We also report parameter counts, training time, and average synthesis time per frame in Table \ref{tab:Parameters}. The results show that the parameter count for \textbf{DM}(\(10.4\)M) is much less than \textbf{SG}(\(109.34\)M). Our method achieves less training time(\(8.02\) min.) than \textbf{SG}(\(6.2 \) hour). However, with the inherent characteristics of diffusion models\cite{yangDiffusionModelsComprehensive2022}, the generation phase in \textbf{DM}(\(1.60\)±\(0.06\)) takes a longer time than SG(\(0.08\)±\(0.06\)).

\subsection{Subjective Evaluation}

\begin{figure}[htp]
\includegraphics[width=0.6\textwidth]{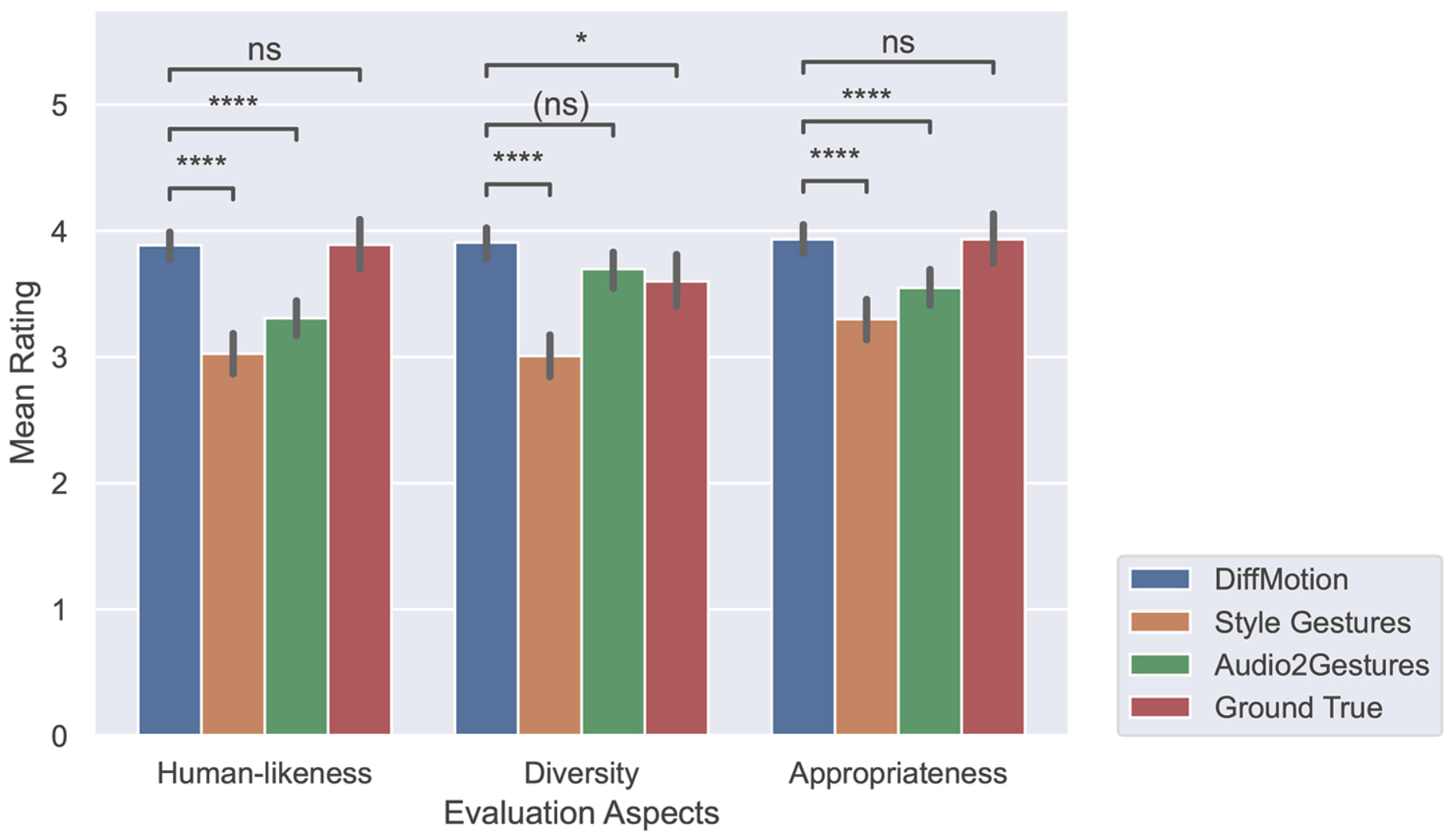}
\caption{Mean ratings with 95\% confidence intervals. Asterisks indicated significant effects(*: \emph{p} \(<\) 0.05, ****: \emph{p} \(<\) 1.00e-04, ns: no significant difference).}\label{fig:meanrating}
\end{figure}

\floatsetup[table]{capposition=top}
\begin{table}[htp]
\caption{The mean perceptual rating score.}\label{tab:overview}
\renewcommand{\belowrulesep}{0pt} 
\renewcommand{\aboverulesep}{0pt}
\begin{tabular}{@{}c|c|c|c@{}}
\toprule
Model & Human-likeness↑                     & Diverse↑           & Appropriateness↑   \\ \midrule
DM    & \textbf{3.89±0.95}                  & \textbf{3.91±0.96} & \textbf{3.93±0.93} \\
SG    & 3.03±1.34                           & 3.01±1.43          & 3.30±1.36          \\
AG    & 3.61±1.12                           & \textbf{3.89±0.92} & 3.55±1.12          \\
GT    & \textbf{3.89±0.93} & 3.60±1.02          & \textbf{3.93±0.92} \\ \bottomrule
\end{tabular}
\end{table}

The ultimate goal of speech-driven gesture generation is to produce natural, convincing motions. Considering that \emph{objective evaluation of gesture synthesis is generally tricky and does not always translate into superior subjective quality for human observes} \cite{alexanderson_simon_style-controllable_2020,wolfertReviewEvaluationPractices2022}, we perform subjective human perception. A question set consisting of three evaluation aspects with a 5-point Likert scale to subjectively evaluate baselines(\textbf{SG}, \textbf{AG}), \textbf{DM}, and \textbf{GT}. Three aspects are \emph{human-likeness}, \emph{diversity}, and \emph{appropriateness}, respectively: 1) \textbf{Human-likeness}: whether the generated gestures are natural and look like the motion of an actual human, without accounting for the speech; 2) \textbf{Diversity}: which motion has more gesture patterns; 3) \textbf{Appropriateness}: the time consistency, that is, whether the generated gestures match the rhythm of the speech.

First, we trained each model and generated 20 clips with the same speech audio as input. Each clip lasts for 18 seconds. Next, we randomly selected 3 clips generated by each model for valuation. Then, we built up the video by GENEA\_visualizer\cite{kucherenko2021large} for user study.

30 volunteer participants were recruited, including 16 males and 14 females, aged 19-23. All of them(20 from China, 10 international students from USA, UK, etc.) are good at English. They were asked to rate the scale for the evaluation aspects. The scores were assigned from 1 to 5, representing the worst to best.

Firstly, we introduced the method to all participants and showed them some example clips which not in the valuation set. After the participants fully understood the process, we started the formal experiment. All participants were instructed to wear headphones and sit in front of a computer screen. The environment was quiet and had no interference.  Participants were unaware of which method each video belonged to. The order of videos was random, but each video was guaranteed to appear three times, presented and scored by the participants. 

One-way ANOVA was conducted to determine if the models' scores differed on the three evaluation aspects. The results are shown in Fig.\ref{fig:meanrating} and Table \ref{tab:overview}. The mean rating scores of \textbf{DM} we proposed are statistically significantly different from the other two baseline models and not statistically significant with \textbf{GT}. The score of human-likeness for \textbf{DM} and \textbf{GT} is \(3.89\)±\(0.95\) and \(3.89\)±\(0.93\), and the appropriateness is \(3.93\)±\(0.93\) and \(3.93\)±\(0.92\), respectively. Interestingly, there was no significant difference between \textbf{DM} (\(3.91\)±\(0.96\)) and \textbf{AG}(\(3.89\)±\(0.92\)) on diversity evaluation and obtained higher scores than \textbf{GT}(\(3.60\)±\(1.02\)). These results suggest that \textbf{DM} is as capable as \textbf{AG} of generating more generous gestures that are not present in ground-truth. The results reveal that the proposed method outperforms previous SOTA methods and demonstrate the diffusion-based method benefits speech-driven gesture generation tasks.

\subsection{Ablation Study}
We found that the length of diffusion step \(N\) is a crucial hyperparameter that can affect the quality and effectiveness of gesture generation. Despite larger \(N\) allowing \(x^N\) to be approximately Gaussian\cite{sohl-dickstein_deep_2015}, it results in more time in the inference process. For making a trade-off between generation efficiency and effectiveness, we tune the number of epochs for early stopping(50 epochs) and evaluated \(N =1, 50,100,200,500, ..., 1000, ..., 2500\) while keeping all other hyperparameters unchanged. The results are listed in Table \ref{tab:DiffusionStep}. Faster generation is achieved but causes jittery when \(N < 100\). A continuous motion can be achieved when \(N \geq 100\)(especially \(N=100\), we reach the best result). However, as \(N\) increases, the time consumed increases substantially. Nevertheless \(N>1000\), the model occasionally produces bizarre poses due to the diffusion step destroying the detail of the raw information\cite{zhangDiffusionNormalizingFlow2021}.

\floatsetup[table]{capposition=top}
\begin{table}[htp]\caption{Evaluation for the number of diffusion step \(N\).}\label{tab:DiffusionStep}
\renewcommand{\belowrulesep}{0pt} 
\renewcommand{\aboverulesep}{0pt}
\begin{tabular}{@{}c|c|c|c|c|c|c|c|c|c|c@{}}
\toprule
Metric                                                                 & 1                                                      & 100                                                  & 300                                                  & 500                                                  & 600                                                  & 800                                                  & 1000                                                 & 1500                                                 & 2000                                                 & 2500                                                 \\ \midrule
\begin{tabular}[c]{@{}c@{}}Synth. time \\ per frame(sec.)\end{tabular} & \begin{tabular}[c]{@{}c@{}}0.004\\ ±0.001\end{tabular} & \begin{tabular}[c]{@{}c@{}}0.34\\ ±0.02\end{tabular} & \begin{tabular}[c]{@{}c@{}}0.97\\ ±0.04\end{tabular} & \begin{tabular}[c]{@{}c@{}}1.60\\ ±0.06\end{tabular} & \begin{tabular}[c]{@{}c@{}}1.91\\ ±0.06\end{tabular} & \begin{tabular}[c]{@{}c@{}}2.55\\ ±0.08\end{tabular} & \begin{tabular}[c]{@{}c@{}}3.23\\ ±0.10\end{tabular} & \begin{tabular}[c]{@{}c@{}}4.81\\ ±0.14\end{tabular} & \begin{tabular}[c]{@{}c@{}}6.43\\ ±0.19\end{tabular} & \begin{tabular}[c]{@{}c@{}}8.02\\ ±0.23\end{tabular} \\
\begin{tabular}[c]{@{}c@{}}Training Time\\ (min.)\end{tabular}         & \begin{tabular}[c]{@{}c@{}}8.63\\ ±0.06\end{tabular}   & \begin{tabular}[c]{@{}c@{}}8.63\\ ±0.07\end{tabular} & \begin{tabular}[c]{@{}c@{}}8.62\\ ±0.10\end{tabular} & \begin{tabular}[c]{@{}c@{}}8.51\\ ±0.08\end{tabular} & \begin{tabular}[c]{@{}c@{}}8.55\\ ±0.09\end{tabular} & \begin{tabular}[c]{@{}c@{}}8.52\\ ±0.08\end{tabular} & \begin{tabular}[c]{@{}c@{}}8.93\\ ±0.08\end{tabular} & \begin{tabular}[c]{@{}c@{}}8.84\\ ±0.08\end{tabular} & \begin{tabular}[c]{@{}c@{}}8.74\\ ±0.07\end{tabular} & \begin{tabular}[c]{@{}c@{}}8.67\\ ±0.08\end{tabular} \\
Train Loss                                                             & 0.99                                                   & 0.32                                                 & 0.20                                                 & 0.17                                                 & 0.16                                                 & 0.15                                                 & 0.13                                                 & 0.12                                                 & 0.10                                                 & 0.09                                                 \\
Val Loss                                                               & 0.99                                                   & 0.34                                                 & 0.22                                                 & 0.19                                                 & 0.18                                                 & 0.16                                                 & 0.15                                                 & 0.13                                                 & 0.11                                                 & 0.10                                                 \\ \bottomrule
\end{tabular}
\end{table}

\section{Conclusion}
In this paper, we propose a novel framework \textbf{DiffMotion} for automatically co-speech gesture synthesis. The framework consists of an Autoregressive Temporal Encoder and a Denoising Diffusion Probability Module. The architecture can learn to obtain the complex one-many mapping between gestures and the accompanying speech and can generate 3D co-speech gesticulation that is natural, diverse, and well-timed. Experiments confirm that our system outperforms previous baselines, quantitatively and qualitatively. Still, there are some limitations. For example, the model tends to generate redundant movements that lack relaxation, and it experiences slow inference due to the inherent characteristics of DDPM. For future work, we may plan the following attempts: 1) considering the breathing space for relaxation; 2) introducing multimodal features such as semantic and affective expressions; 3) investigating a real-time system enabling users to interact with virtual human interfaces.   

\subsubsection{Acknowledgements.} This work was supported by the Key Program and development projects of Zhejiang Province of China (No.2021C03137), the Public Welfare Technology Application Research Project of Zhejiang Province, China (No.LGF22F020008), and the Key Lab of Film and TV Media Technology of Zhejiang Province (No.2020E10015).

%
%
%
\bibliographystyle{splncs04}
\bibliography{mybibliography}

\end{document}